\documentstyle[12pt,epsfig]{article}
\newcommand{\beq} { \begin{eqnarray} }
\newcommand{\eeq} { \end{eqnarray} }
\def\bea{\begin{eqnarray}}
\def\eea{\end{eqnarray}}
\def\bef{\begin{figure}}
\def\eef{\end{figure}}
\def\bet{\begin{table}}
\def\eet{\end{table}}%
\def\eq#1{{eq.~(\ref{#1})}}

\def\fig#1{{Fig.~\ref{#1}}}
\def\figz#1#2{{Fig.~\ref{#1}{#2 }}}
\def\tab#1{{Table~\ref{#1}}}
\def\ra{\rightarrow}
\def\ot{\otimes}

\def\lsim{\raise0.3ex\hbox{$\;<$\kern-0.75em\raise-1.1ex\hbox{$\sim\;$}}}
\def\gsim{\raise0.3ex\hbox{$\;>$\kern-0.75em\raise-1.1ex\hbox{$\sim\;$}}}
\def\VEV#1{\left\langle #1\right\rangle}
\def\21{$SU(2) \ot U(1)$}
\def\321{$SU(3) \ot SU(2) \ot U(1)$}

\newcommand{\chim} [1] {\tilde{\chi}^{-}_{#1} }
\newcommand{\chimp} [1] {\tilde{\chi}^{\mp}_{#1} }
\newcommand{\chip} [1] {\tilde{\chi}^{+}_{#1} }
\newcommand{\chipm} [1] {\tilde{\chi}^{\pm}_{#1} }
\newcommand{\chiz} [1] {\tilde{\chi}^{0}_{#1} }

\newcommand{\tanb} {\tan \beta}

\newcommand {\pp} {$p \bar{p}$ and $p p$ }
\newcommand {\glu} {\tilde{g} }
\newcommand {\mglu} {m_{\tilde{g}} }

\def\nps#1#2#3{          {\it Nucl. Phys. B (Proc. Suppl.) }
                         {\bf #1} (19#2) #3} 
\def\np#1#2#3{           {\it Nucl. Phys. }{\bf #1} (19#2) #3}
\def\pl#1#2#3{           {\it Phys. Lett. }{\bf #1} (19#2) #3}
\def\pr#1#2#3{           {\it Phys. Rev. }{\bf #1} (19#2) #3}
\def\prep#1#2#3{         {\it Phys. Rep. }{\bf #1} (19#2) #3}
\def\prl#1#2#3{          {\it Phys. Rev. Lett. }{\bf #1} (19#2) #3}

\def\n.c.#1#2#3{         {\it Nuovo Cim. }{\bf #1} (19#2) #3}
\def\r.n.c.#1#2#3{       {\it Riv. del Nuovo Cim. }{\bf #1} (19#2) #3}

\relax
\begin{document}
\thispagestyle{empty}
\begin{titlepage}
\begin{center}
\rightline{hep-ph/9612436}
\hfill \today \\
\vskip 0.2cm
{\Large \bf Signatures of Spontaneous Breaking of R-parity in Gluino Cascade
       Decays at LHC}
\vskip 0.1cm
{\large A. Bartl}$^*$, 
{\large W. Porod}$^{\dag}$ \\
{\it Institut f\"ur Theoretische Physik, University of Vienna \\
A-1090 Vienna, Austria}\\
\vskip 0.1cm
{\large F. de Campos}$^{\ddag}$ \\
{\it Instituto de F\'{\i}sica Te\'orica - Universidade Estadual Paulista \\
     Rua Pamplona, 145 - 01405-900 - S\~ao Paulo - SP, Brasil } \\
\vskip 0.1cm
{\large M. A. Garc\'{\i}a-Jare\~no}$^{\S}$,
{\large M. B. Magro}$^{\P}$ and
{\large J. W. F. Valle}$^{\diamond}$\\
{\it Instituto de F\'{\i}sica Corpuscular - C.S.I.C.\\
Departament de F\'{\i}sica Te\`orica, Universitat de Val\`encia\\
46100 Burjassot, Val\`encia, Spain\\
 http://neutrinos.uv.es}\\
\vskip .1cm
{\large W. Majerotto}$^{\flat}$\\
{\it Institut f\"ur Hochenergiephysik, Akademie der Wissenschaften \\
A-1050 Vienna, Austria}\\
\vskip 0.1cm
{\bf Abstract}
\end{center}
\begin{quotation}
\noindent
We study the pattern of gluino cascade decays in a class of supersymmetric 
models where R-parity is spontaneously broken.
We give a detailed discussion of the R-parity 
violating decays of the lightest neutralino, the second lightest 
neutralino and the lightest chargino. The multi-lepton 
and same-sign dilepton signal rates expected in these models are compared with 
those predicted in the Minimal Supersymmetric Standard Model.
We show that these rates can be substantially enhanced in 
broken R-parity models.

\end{quotation}
\end{titlepage}
\section{Introduction}

The search for supersymmetry (SUSY) will play an important r\^ole
in the experimental program of LHC which will explore the mass range of
supersymmetric particles up to TeV energies \cite{aachen}.
Due to the high production cross section of gluinos and squarks at 
hadron colliders, their signals are expected to be well above the 
Standard Model (SM) background.

Up to now most studies of gluino production and decays 
\cite{aachen,barnett} have been carried out within
the framework of the {\sl Minimal Supersymmetric Standard Model (MSSM)}
\cite{mssm}. In the MSSM R-parity is conserved implying that the lightest 
supersymmetric particle (LSP) is stable, giving rise to the 
missing energy (\mbox{$E_T \hspace{-4.5mm}/$}\,) signal. There are already some
studies where the effects of R-parity breaking have been explored
\cite{binet}. However, most of them are in the context of models 
with explicit breaking of R-parity.

Although the MSSM is by far the most
well studied realization of supersymmetry, there is considerable 
theoretical as well as phenomenological interest in studying the
implications of alternative scenarios without 
 R-parity conservation \cite{beyond}. The violation of 
R-parity could arise  explicitly \cite{expl} as a residual 
effect of some larger unified theory \cite{expl0},
or spontaneously, through nonzero vacuum expectation values (VEV's)
for scalar neutrinos \cite{aul}. In the first case there is a large number
of unknown parameters characterizing the superpotential of these models. 
For simplicity these effects are usually studied assuming that there 
is only one dominant term which breaks R-parity explicitly. In contrast,
models with spontaneous R-parity breaking \cite{MASIpot3,MASI,ROMA,ZR} 
depend on much fewer parameters, which allow a more systematic
study of  R-parity breaking signals. 
Moreover, in these models the scale of R-parity violation 
is expected to lie in the TeV range \cite{MASIpot3}. This leads 
to effects that can 
be large enough to be experimentally observable, for a wide
range of parameter choices  consistent with observations,
including astrophysics and cosmology \cite{beyond,ROMA,ROMA1}.

There are two generic cases of spontaneous R-parity breaking models.
If lepton number is part of the gauge symmetry there is a new
gauge boson $Z^\prime$ which gets mass via the Higgs mechanism
\cite{ZR}. In this model
the lightest SUSY particle (LSP) is in general a neutralino which
decays mostly into visible states, therefore breaking R-parity. 
The main decay modes are decays such as
\beq
\label{vis}
\chiz{1} \ra Z^{(*)} \nu \ra f \bar{f} \nu,
\eeq
where the $Z$-boson can be either virtual or real and f denotes a charged 
fermion. Its invisible decay modes
are in the channels $\chiz{1} \ra 3 \nu$.
Alternatively, if spontaneous R-parity violation occurs
in the absence of any additional gauge symmetry, it leads to the
existence of a physical massless Nambu-Goldstone boson, called
majoron (J). In this case {\sl the lightest SUSY
particle is the majoron} which is massless and therefore stable
\footnote{The majoron may have a small mass due to explicit
breaking effects at the Planck scale. In this case it may decay
into neutrinos and photons. However, the time scales are only of
cosmological interest and do not change the signal expected in
laboratory experiments \cite{KEV}.}. As a consequence the lightest
neutralino $\chiz{1}$ may decay invisibly as
\beq
\label{invis}
\chiz{1} \ra \nu + J.
\eeq
This decay conserves R-parity since the majoron has a large
R-odd singlet sneutrino component \cite{MASIpot3,MASI}.
 
We also consider a specific class of models
with explicit R-parity breaking characterized by a single 
bilinear superpotential term of the type $\ell H_u$
\cite{epsi}. These models mimic in many 
respects the features of models with spontaneous breaking 
of R-parity containing an additional gauge boson. 
Since they do not contain the majoron, the lightest
neutralino has only decays into Standard Model fermions.
In the following the class of models containing a majoron will 
be denoted by the majoron-model, whereas the models without a 
majoron will be denoted generically by the $\epsilon$-model \cite{epsi}.

Although in these broken R-parity models supersymmetric particles may be 
produced singly (see ref. \cite{bounds}), it is most likely that gluinos 
at the LHC will be produced in pairs, and that R-parity violation
will affect only the structure of their cascade decays. The
most obvious modification of these cascade decays with respect 
to the one expected in the MSSM comes from the fact that the 
lightest neutralino now can decay.

In this paper we concentrate on cascade
decays of the gluino assuming that it is lighter than the
squarks. We pay special attention to the impact of R-parity violation 
in the the multi-lepton (ML) signals and the same-sign dilepton (SSD) 
signals.The gluino has the following R-parity conserving decays:
\beq
\label{gdecay}
\glu & \to & q\bar{q}\chiz{i}\;;\; q\bar{q}'\chipm{j}\;;\;g \chiz{i}
\eeq
where $\chiz{i}$ denotes the neutralinos and $\chipm{j}$ the charginos.
In \eq{gdecay} we also include the decays into top quarks, which give important
contributions to the ML and SSD signals.
If R-parity is violated spontaneously one has in principle also the
decay modes $\glu \to  q + \bar{q}' + l, \,  q + \bar{q} + \nu_l$. 
We have neglected these decays in the 
following because their branching ratios are expected to be much smaller. 

In order to characterize the complicated pattern of gluino 
cascade decays in broken R-parity models, we will first discuss
the decays of the gluinos into charginos and neutralinos (Section 2).
Then we will discuss the decay pattern of
the lightest neutralino, the second lightest neutralino and the 
lightest chargino, paying particular attention to
R-parity violating decays (Section 4). We calculate the rates for
the 3-, 4-, 5- and 6-lepton signals and the same-sign 
dilepton signal in the two classes of broken R-parity models described above
and compare them with the corresponding rates predicted in the MSSM 
(Section 5).

\section{Gluino decays into charginos and neutralinos}

At \pp colliders gluinos are produced through $g g$ and  $q \bar{q}$ fusion
\cite{Dawson85}.  Here we
will assume that squarks are heavier than gluinos, so that pair
production of gluinos dominates. As an example we will consider a  
gluino with a mass of 500 GeV. At  the LHC with a centre of mass energy 
of 14 TeV the production cross section will be $\sim 25$ pb, which 
corresponds to 2.5 million gluino-pairs per year for an integrated 
luminosity of $10^5 pb^{-1}$.

As the multi-lepton signal is the result of a complicated decay chain,
one has to calculate the branching ratio of each step in the gluino cascade 
decays. For the computation of the decays in \eq{gdecay} we have used the 
formulae given  in
\cite{Bartl91}. As these formulae have been developed in
the framework of the MSSM, one has to make appropriate replacements as 
given in  Appendix~A in order to be consistent with the models described 
in the next section.

We will consider a low $\tanb$
scenario ($\tanb=2$) and a high $\tanb$ scenario ($\tanb=30$). These
choices are theoretically motivated by renormalization group
studies in some unified supergravity models \cite{btop}.
Moreover we will vary $\mu$ between $-$1000~GeV and 1000~GeV.
In \fig{glubranch} we show the gluino decay branching ratios.
We have summed over all quark flavours and included the contribution 
coming from the decay 
into a gluon and a neutralino.
As already mentioned, the R-parity violating decays of the gluino can
be neglected.
Because of kinematics (the masses of the heavier neutralinos $\chiz{3}$ and 
$\chiz{4}$ and the heavier chargino $\chim{2}$ is of order $\mu$ if 
$|\mu| > M_2$)  for $|\mu| > \mglu / 2$ we have only decays into the 
lightest chargino $\chim{1}$ ($\sim 50\%$) and the two lightest neutralinos 
$\chiz{1}$ ($\sim 20\%$) and $\chiz{2}$ ($\sim 30\%$).
For $|\mu| < \mglu/2$ the 
decay into the heavier chargino becomes important ($\gsim 25\%$).

The charginos and neutralinos arising from gluino decays will 
subsequently decay as discussed above, leading to the various 
multi-lepton signals we will discuss in Section~\ref{sec:discus}. 
Another important source of leptons are the top quarks
produced in $\glu \to t b \chipm{j}$ with the top quark decaying into a
$W$-boson. The branching ratio of this decay is at least 5\%.

For the case $\tanb = 30$ there are some changes in the 
area $|\mu| < \mglu /2$
compared to the case $\tanb = 2$. Because of larger bottom
Yukawa couplings the decays $\tilde g \to b \bar{b} \chiz{i}$ are enhanced. 
However, for the multi-lepton signal these changes are only important for 
a parameter region which is already excluded by experimental data. For
further details about gluino decays see ref. \cite{Bartl91}.

\section{Lepton-Gaugino-Higgsino Mixing}

The basic tools in our subsequent discussion are the chargino
and neutralino mass matrices. The chargino mass matrix may be written as
\cite{MASIpot3}
\beq
\begin{array}{c|cccccccc}
& e^+_j & \tilde{H^+_u} & -i \tilde{W^+}\\
\hline
e_i & h_{e ij} v_d & - h_{\nu ij} v_{Rj} & \sqrt{2} g_2 v_{Li} \\
\tilde{H^-_d} & - h_{e ij} v_{Li} & \mu & \sqrt{2} g_2 v_d\\
-i \tilde{W^-} & 0 & \sqrt{2} g_2 v_u & M_2
\end{array}
\label{chino}
\eeq
Its diagonalization requires two unitary matrices U and V 
\bea
{\chi}_i^+ = V_{ij} {\psi}_j^+\\
{\chi}_i^- = U_{ij} {\psi}_j^- ,
\label{INO}
\eea
where the indices $i$ and $j$ run from $1$ to $5$ and
$\psi_j^+ = (e_1^+, e_2^+ , e_3^+ ,\tilde{H^+_u}, -i \tilde{W^+}$)
and $\psi_j^- = (e_1^-, e_2^- , e_3^-, \tilde{H^-_d}, -i \tilde{W^-}$).

The details of the neutralino mass matrix are
rather model dependent. However, for our purposes it will be sufficient
to use the following effective form given in \cite{MASIpot3}
\beq
\begin{array}{c|cccccccc}
& {\nu}_i & \tilde{H}_u & \tilde{H}_d & -i \tilde{W}_3 & -i \tilde{B}\\
\hline
{\nu}_i & 0 & h_{\nu ij} v_{Rj} & 0 & g_2 v_{Li} & -g_1 v_{Li}\\
\tilde{H}_u & h_{\nu ij} v_{Rj} & 0 & - \mu & -g_2 v_u & g_1 v_u\\
\tilde{H}_d & 0 & - \mu & 0 & g_2 v_d & -g_1 v_d\\
-i \tilde{W}_3 & g_2 v_{Li} & -g_2 v_u & g_2 v_d & M_2 & 0\\
-i \tilde{B} & -g_1 v_{Li} & g_1 v_u & -g_1 v_d & 0 & M_1
\end{array}
\label{nino}
\eeq
This matrix is diagonalised by a $7 \times 7$ unitary matrix N,
\beq
{\chi}_i^0 = N_{ij} {\psi}_j^0 ,
\eeq
where
$\psi_j^0 = ({\nu}_i,\tilde{H}_u,\tilde{H}_d,-i \tilde{W}_3,-i \tilde{B}$),
with $\nu_i$ denoting the three weak-eigenstate neutrinos.

In the above  equations  $v_R$ is the VEV of the right sneutrino mostly
responsible for the spontaneous violation of R-parity
\footnote{There is also a small seed of R-parity
breaking in the doublet sector, $v_L = \VEV {\tilde{\nu}_{L\tau}}$,
whose magnitude is related to the Yukawa coupling
$h_{\nu}$. Since this vanishes as $h_{\nu} \ra 0$,
we can naturally obey the limits from stellar energy
loss \cite{KIM}}. 
The VEV's $v_u$ and $v_d$ are the usual ones responsible
for the breaking of the electroweak symmetry and the generation of
fermion masses, with the combination $v^2 = v_u^2 + v_d^2$
 fixed by the $W,Z$ masses.
Moreover, $M_{1,2}$ denote the supersymmetry
breaking gaugino mass parameters and $g_{1,2}$ are the
\21 gauge couplings divided by $\sqrt{2}$. In the following we assume the
GUT relation $\frac{3}{5} M_1/M_2 = \tan^2{\theta_W}$.
Note that the effective Higgsino mixing parameter $\mu$ may be
given in some models as $\mu = h_0 \VEV \Phi$, where $\VEV \Phi$
is the VEV of an appropriate singlet scalar.
In the $\epsilon$-model the term $h_{\nu} v_R$ in \eq{chino} and (\ref{nino})
is replaced by a mass parameter $\epsilon$ \cite{epsi}.

There are restrictions on these parameters that follow 
from searches for SUSY particles at LEP \cite{LEPSEARCH,magro} and 
at TEVATRON \cite{D0}. In addition, we take into
account the constraints from  neutrino physics and 
weak interactions phenomenology \cite{bounds},
which are more characteristic of R-parity breaking models.
These are important, as they exclude 
many parameter choices that are otherwise allowed by the 
constraints from the collider data, while the converse is 
not true. Due to these constraints R-parity violation effects 
manifest themselves mainly in the third generation. We therefore
assume $v_{L_1} = v_{L_2} = v_{R_1} = v_{R_2} =  0$.

Most of our subsequent analysis will be general enough to
cover a wide class of \21 models with spontaneously broken R-parity,
such as those of ref. \cite{MASIpot3,MASI}, as well as models
where the majoron is absent due to an enlarged gauge structure
\cite{ZR}. Many of the phenomenological features relevant
for the LHC studies discussed here are already present in the 
$\epsilon$-model which effectively mimics the spontaneous violation
of R-parity through an explicit R-parity-breaking bilinear superpotential 
term $\ell H_u$ \cite{epsi}. 

In the following we will need the mass eigenstates of (\ref{chino}) and
(\ref{nino}). In order to use the same notation as in the
MSSM, $l$, $\nu_l$ denote the mass eigenstates for charged leptons
and neutrinos, $\chiz{i}$ (i=1,..,4) the mass eigenstates for neutralinos
and $\chipm{j}$ (j=1,2) the mass eigenstates for charginos.

\section{Neutralino and Chargino Decays}
\label{sec:chaneu}

In this section we shall discuss in detail
the decays of charginos and neutralinos which occur in the cascade decays
of the gluino. The neutralinos have the following R-parity conserving two-body 
decay modes:
\beq 
\label{neutdec2}
\chiz{i} & \to & \chipm{j} W^{\mp}\;;\; \chiz{k} Z. 
\eeq
One has in principle also decays into Higgs bosons, squarks and
sleptons which we neglect. Insofar as squarks and sleptons are concerned, 
we simply assume them to be too heavy to be important. A notable exception
for the case of the {\sl majoron model} considered in this paper is that 
of the majoron, which is a linear combination of the \21 singlet
sneutrinos and is massless (or very light) because it is
a Goldstone boson. Indeed, in this case the decay $\chipm{j} \ra
\tau^\pm + J$ can have a sizeable branching ratio. We have taken
into account the existence of such decays in the evaluation
of the multi-lepton (ML) and same-sign dilepton (SSD) rates
presented in this paper. We have, however, not studied the
corresponding signals for the LHC, because they involve the 
detection of taus in the final state. For the case of LEP2
these signals have been already considered in the literature 
\cite{tauJ0}, although more work is needed \cite{tauJ}.

Since R-parity is violated one has the additional decay modes:
\beq 
\chiz{i} & \to & l^{\pm}_j W^{\mp}\;;\;\nu_l Z 
\eeq
In case these two-body decays into gauge bosons are kinematically 
forbidden the neutralinos have the following three-body decay modes:
\beq
\label{neutdec}
\chiz{i} & \to & \chiz{j} f \bar{f} \;;\;
\chipm{j} f' \bar{f} \;;\;
\nu_k f \bar{f} \;;\;
l^{\pm}_k f' \bar{f}
\eeq
where $f,f' = l_i,\nu_i,d_i,u_i$.
The first two decays conserve R-parity whereas the other ones violate 
R-parity. 
In addition, in \21 models with spontaneous R-parity violation one has also 
the decay into a majoron $J$ 
\beq
\label{majdecay}
\chiz{i} \to \chiz{j} J, \nu_k J. 
\eeq
It is important to notice that the decays into
a standard neutrino conserve R-parity, since
the majoron is mainly a right sneutrino, and thus R-odd.

Turning now to charginos, the lightest one has the following 
R-parity conserving two-body decay:
\beq
\label{char2b2}
\chip{1} \to\chiz{i} W^+ 
\eeq
Again we assume that decays into scalars
are kinematically forbidden. In addition it has the following 
R-parity-violating decay modes:
\beq
\label{char2b}
\chip{1} & \to & \nu_j W^+ \;;\l^+_j Z 
\eeq
In models with majoron the chargino can decay according to
\cite{ROMA}
\beq
\label{charmaj}
\chip{1} \to l_j J 
\eeq

In case two-body decays into gauge bosons are 
kinematically forbidden, the chargino decays as:
\beq  
\label{chardecay2}
\chip{1} & \to & \chiz{j} f' \bar{f}\;;\;
\nu_k f' \bar{f} \;;\;
l^{+}_k f \bar{f}
\eeq
where the first decay conserves and the others
break R-parity. The formulae for two- and three-body decay widths
relevant for our study are given in Appendix~B.

As already mentioned in the previous section R-parity violating effects
manifest themselves mainly through the mixing of the third generation
leptons with charginos and neutralinos. 
In order to show typical examples of neutralino and chargino decays
we have fixed the following set of parameters:
$M_2 = 170$~GeV, $m_A = 500$~GeV,
$h_{\nu 33} = 0.01$, $v_R \equiv v_{R3} = 100$~GeV and
$v_L \equiv v_{L 3} = 10^{-5}$~GeV. In the $\epsilon$-model
this corresponds to $\epsilon = 1$~GeV. As we already mentioned,
we have considered $\tanb$ in both low and high value scenarios,
$\tanb=2$ and $\tanb=30$, as suggested by renormalization group
studies \cite{btop}. The $\mu$ parameter has been varied between 
$-$1000~GeV and 1000~GeV.

In contrast to the MSSM, in models where R-parity is broken the lightest 
neutralino will decay. Let us first focus on the majoron-model. 
In \fig{brchi01m} we show the branching ratios for the lightest 
neutralino for $\tan{\beta}=2$. 
The decay into the majoron dominates for two reasons: first, the 
decay of the lightest neutralino into a neutrino and a 
majoron is R-parity conserving, while the decays into $W$ and $Z$
bosons are R-parity violating. Moreover, the decays into gauge-bosons 
are either phase space suppressed two-body decays or three-body decays.
Note that the decays
into a majoron and a neutrino and into three neutrinos are
invisible decays thus leading to the same missing transverse
momentum signature characteristic of a stable neutralino in 
the MSSM. The importance of the decays into the majoron 
increases for larger $\tan \beta$. 
We found that in the case of $\tanb = 30$
the decay into the majoron is practically 100~\% in the parameter
range which will not be covered by LEP2.

Let us now turn to the $\epsilon$-model. In \figz{brchi01e} {a (b)} 
we present the branching ratios for the lightest neutralino for 
$\tan{\beta}=2$ (30). We can see that for most $\mu$ values the $W$ 
channel dominates over $Z$ channel. The reason for this behaviour is 
the fact that for our parameter choices, very often the neutralino has 
charged-current two-body decays and neutral current three-body decays.
Another important fact is that the $Z$-boson only couples to 
the Higgsino components of the neutralino which are rather small in our case.

In the MSSM, the second lightest neutralino 
will mainly decay into a $Z$-boson and the lightest neutralino due to 
kinematics. In \fig{brchi02m} we show the branching ratios for the second
lightest neutralino for $\tan{\beta}=2$ in the majoron-model. Notice
that, the $\nu Z$ and $\tau W$ decay 
modes are  sizeable, even though they violate  R-parity, since they are 
kinematically favoured with respect to the 
R-parity conserving MSSM decay mode $\chiz{1} Z^*$ (the case where the second 
lightest neutralino is lighter than the $W$-boson will be 
completely covered by LEP2, for our choice of parameters). On the other hand
the $\chiz{2} \to J \nu_{\tau}$ channel, although R-parity 
conserving, is smaller since the underlying Yukawa coupling is relatively 
small ($10^{-2}$). This is in sharp contrast to the situation in $\chiz{1}$ 
decay 
(\fig{brchi01m}) where the charged and neutral current induced
decays are suppressed by phase space. Note also that here we do 
not show the decay $\chiz{2} \to \chim{1} W$, because it is only 
important in a small range which will be covered by LEP2. For 
$\tan{\beta}=30$ the R-parity breaking decays are negligible.
For the case of the $\epsilon$-model the R-parity violating decays 
into gauge-bosons are again significant if $\tanb$ is small, as 
can be seen in \fig{brchi02e} (again we do not show the decay 
$\chiz{2} \to \chim{1} W$). For $\tanb = 30$ they become negligible.

In the MSSM the lightest chargino decays mainly into the lightest 
neutralino and a $W$-boson, since the decays into the second lightest 
neutralino are suppressed by phase space. In \figz{brchipm}{a (b)} we 
show the branching ratios in the majoron-model of the lightest chargino 
as a function of $\mu$ for $M_2=170$~GeV and $\tanb = 2(30)$. 
In contrast, for the case of the $\epsilon$-model all decays of 
the lightest chargino are induced by W and Z boson exchange.
For the ranges of $M_2$ and $\mu$ considered in this paper,
all of these decays are two-body. As a result the R-parity 
violating decay branching ratios are all negligible.

Although our discussion has been quite general, we have neglected, 
as already mentioned, chargino and neutralino decays mediated by scalar
particles, including Higgs bosons.
In this approximation, neutralinos and charginos produced by gluino 
decays have only decays mediated by $W$ or $Z$ bosons, except for the 
two-body majoron decays, characteristic of the spontaneous 
R-parity breaking models with the minimum \21 gauge symmetry.
 
\section{Multi-lepton and same-sign dilepton rates}
\label{sec:discus}

In the following we calculate the multi-lepton (ML) and same-sign 
dilepton (SSD) rates in gluino pair 
production for the MSSM, the majoron-model and the $\epsilon$-model. 
We have counted all leptons coming from charginos, neutralinos, 
$t$-quarks, $W$- and $Z$-bosons, summing over electrons and muons. We again 
take $\mglu = 500$~GeV, and all other parameters as in 
Section~\ref{sec:chaneu}.

Quite generally, the various ML rates in the R-parity violating models
can be different from those in the MSSM for two reasons:
(i) The lightest neutralino $\chiz{1}$ can decay leptonicaly 
as $\chiz{1} \to Z^{(*)} \nu_{\tau} \to l^+ l^- \nu_{\tau}$, 
$\chiz{1} \to W^{(*)} \tau \to l^+ \nu_l \tau$, leading to an 
enhancement of the multi-lepton rates. 
(ii) The R-parity violating decays of the lightest chargino
$\chimp{1}$ and the second lightest neutralino $\chiz{2}$ may 
reduce the leptonic signal, $\chiz{2} \to W^{(*)} \tau$, 
$J \nu_{\tau}$, $\chim{1} \to J \tau$.
Depending on which of these two effects is dominant, one has an
overall enhancement or a reduction of the leptonic rates compared 
to those expected in the MSSM. A summary of the effects of the most 
important R-parity breaking decay modes is given in \tab{tab:decay}.

In \fig{multilep02} we show the branching ratios for the 3-, 4-, 5- and 
6-lepton events for $\tanb = 2$. In comparison with the MSSM, the 
majoron-model exhibits the feature that the overall rates for the
ML signals are enhanced for $\mu < 0$ and suppressed for $\mu > 0$.
This is due to the fact that the R-parity violating decays of the lightest
neutralino into gauge bosons have a larger branching ratio for $\mu < 0$ than
for $\mu > 0$, and that the R-parity violating decays of the second
lightest neutralino are larger for $\mu > 0$ than for $\mu < 0$.
Note that for $\mu < 0$ the 5-lepton signal is much larger in the 
majoron-model than in the MSSM, giving about 30 to 1200 events per 
year for a luminosity of $10^{5}pb^{-1}$. 
The 6-lepton signal has a rate up to $5 \times 10^{-5}$ in the range
$-300$~GeV$< \mu < -80$~GeV giving 125 events per year.

For the case of the $\epsilon$-model the ML rates are enhanced 
compared to the MSSM and the majoron-model for all $\mu$. The
reason is that the lightest neutralino always decays into
a gauge boson (either real or virtual) which further decays into leptons.
This over-compensates the reduction of leptons coming from the second
lightest neutralino. In this model the 3- and 4-lepton signals are  
enhanced by an order of magnitude compared to the MSSM.
The branching ratio for the 5-lepton signal is larger than $2 \times 10^{-4}$
and the branching ratio for the 6-lepton signal goes up to $5 \times 10^{-4}$.

In \fig{multilep30} we show the ML signal rates for $\tanb = 30$. 
As one can see, for the 6-lepton signal they are larger than for 
$\tanb = 2$. For $|\mu| > 200$ GeV the majoron-model and the MSSM 
give similar results because the lightest  neutralino decays mainly 
invisibly and the R-parity violating decays of the second lightest 
neutralino are somewhat smaller than the
conventional ones. In the $\epsilon$-model again all ML signals are 
enhanced compared to the MSSM. For example the 5-lepton rate is 
larger than $3 \times 10^{-4}$.

In \fig{ssdlep} we show the SSD signal for $\tanb =2$ and 30.
In the case of $\tanb =2$ the signal is enhanced in the majoron-model for  
$\mu \lsim -100$ GeV or $\mu \gsim 200$ GeV. This is due to the fact that
now at least one of the neutralinos has a sizeable branching ratio into a 
$W$, leading to the enhancement of the signal (see \tab{tab:decay}). 
In the $\epsilon$-model the signal is larger by an order of magnitude 
except for $|\mu| \lsim 200$~GeV. In the case of $\tanb = 30$ again 
the majoron-model and the MSSM give similar results whereas in the 
$\epsilon$-model the signal is one order of magnitude larger than in the MSSM.

\section{Conclusions}

We have studied the effects of R-parity violation in gluino cascade
decays for two different classes of models, the majoron-model and the
$\epsilon$-model. We have calculated the rates for the ML and SSD signals. 
These processes are interesting from the experimental point of view since for
example, the 4-, 5-, 6-lepton signal are practically free of 
background from Standard Model processes. 

In order to understand the complex decay pattern
a detailed analysis of the decays of neutralinos and charginos has been
performed. In particular, it has been shown that not only the R-parity
violating decays of the lightest neutralino, but also those of the 
second lightest neutralino and the lightest chargino are important.
Comparing the majoron-model with the MSSM, the ML and SSD signals can
increase or decrease depending on the model parameters. Especially
for small $\tanb$ and  negative $\mu$ the MSSM and the majoron-model give
different results. In the $\epsilon$-model all signals are enhanced
by one order of magnitude for most of the parameter ranges considered.

The results found in this paper should encourage one to perform
detailed Monte Carlo simulations in order to take into account all 
the detector features relevant in an experiment.

{\bf Acknowledgements}                      
                                    
This work was supported by DGICYT grant PB95-1077,
 by EEC under the TMR contract ERBFMRX-CT96-0090, and by
{\sl Fonds zur F\"orderung der wissenschaftlichen Forschung}, 
Projekt Nr. P10843-PHY. In addition, we have been supported
by Acci\'on Integrada Hispano-Austriaca (''B\"uro f\"ur 
technisch-wissen\-schaft\-liche Zusammenarbeit des \"OAD'',
project no. 1); and by Fellowships grants by DGICYT (M. A. G. J.),
FAPESP (M. B. M.), as well as CNPq-Brazil (M. B. M. and F. de C.).

\newpage
\begin{table}
\begin{center}
 \begin{tabular}{lcc} 
  \hline \noalign{\smallskip}
    decay mode & ML signal & SSD signal   \\
  \noalign{\smallskip} \hline \noalign{\smallskip}
   $\chiz{1} \to Z^{(*)} \nu_{\tau}$ & + & 0 \\
   $\chiz{1} \to W^{(*)} \tau$  & + & + \\
   $\chiz{1} \to J \nu_{\tau}$ & 0 &  0 \\
   $\chiz{i} \to Z^{(*)} \nu_{\tau}$ & 0 & 0 \\
   $\chiz{i} \to W^{(*)} \tau$ & - & + \\
   $\chiz{i} \to J \nu_{\tau}$ & - & - \\
   $\chim{1} \to W^{(*)} \nu_{\tau}$ & 0 & 0 \\
   $\chim{1} \to J \tau$ & - &  - \\
    \noalign{\smallskip} \hline 
 \end{tabular}

 \caption[]{\label{tab:decay} 
    Influence of the most important R-parity violating decays on the 
    multi-lepton and same-sign dilepton signals. As reference
    model we take the MSSM. For neutralino decays one has to
    distinguish between the lightest neutralino and the heavier ones.
    We therefore list first the decays of the lightest one and afterwards
    the decays of the heavy ones (i=2,3,4). Here $+$ ($-$) denotes an 
   enhancement (suppression) of the signal with respect to that expected in the MSSM,
  whereas 0 denotes that there is no
    difference compared to the MSSM.}
\end{center}
\end{table}

\newcommand{\recht} {\begin{picture}(7,7)
                      \linethickness{1.7mm}
                      \put(2,1){\line(0,1){5}}
                      \thinlines
                     \end{picture}
                    } 
\newcommand{\rechtl} {\begin{picture}(7,7)
                      \put(2,1){\framebox(5,5){}}
                     \end{picture}
                    } 

\newpage

\noindent
\begin{minipage}[t]{75mm}  
{\setlength{\unitlength}{1mm}
\begin{picture}(75,77)
\put(3,4){\mbox{\psfig{figure=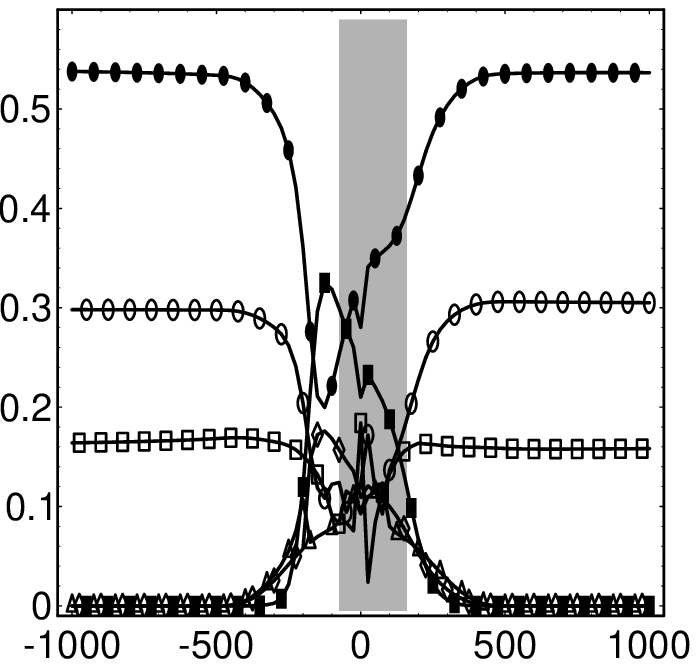,height=6.6cm}}}
\put(8,70){\makebox(0,0)[bl]{{\small $BR(\glu)$}}}
\put(69.5, 0.3){\makebox(0,0)[br]{{\small $\mu$~[GeV]}}}
\end{picture}}
\refstepcounter{figure}
\label{glubranch}
\begin{small}
{\bf \fig{glubranch}}:
The branching ratios for $\glu \to \chiz{i} + q + \bar{q}, \chiz{i} + g$
and $\glu \to \chipm{j} + q + \bar{q}'$ (summed over all quark flavours)
as a function of $\mu$. We have taken $\mglu = 500$~GeV, 
$m_{\tilde q_i} = 2 \, \mglu$, $\tanb = 2$, $m_t = 175$~GeV and $m_b = 5$~GeV.
The curves correspond to the following transitions:
 \rechtl into $\chiz{1}$,$\circ$ into $\chiz{2}$,
 $\triangle$ into $\chiz{3}$, $\diamondsuit$ into $\chiz{4}$, 
 $\bullet$ into $\chimp{1}$ and \recht into $\chimp{2}$.
The  shaded area will be covered by LEP2.
\end{small}
\end{minipage}
\hspace{6mm}
\noindent
\begin{minipage}[t]{75mm}  
{\setlength{\unitlength}{1mm}
\begin{picture}(75,77)
\put(3,4){\mbox{\psfig{figure=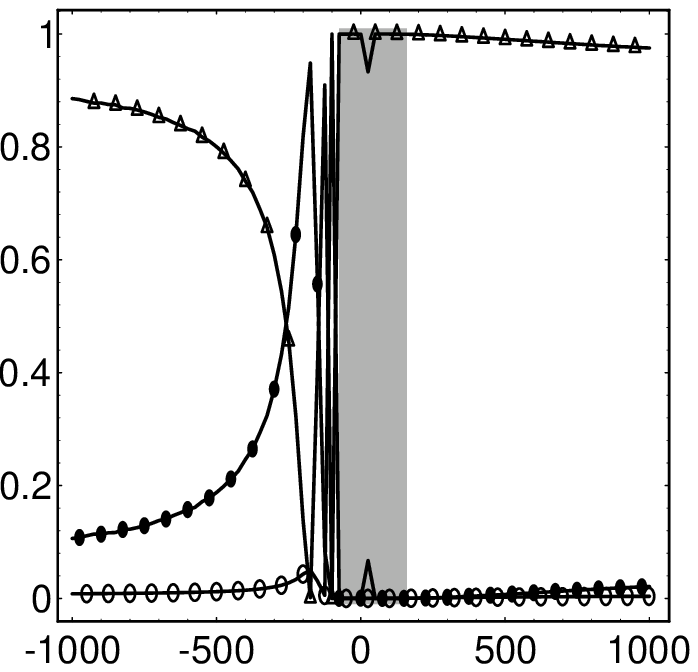,height=6.6cm}}}
\put(8,70){\makebox(0,0)[bl]{{\small $BR(\chiz{1})$}}}
\put(69.5, 0.3){\makebox(0,0)[br]{{\small $\mu$~[GeV]}}}
\end{picture}}
\refstepcounter{figure}
\label{brchi01m}
\begin{small}
{\bf \fig{brchi01m}}:
Branching ratios of the lightest neutralino in the majoron-model. We
have taken $M_2 =170$~GeV, $h_{\nu33} = 0.01$, $v_R = 100$~GeV,
$v_L = 10^{-5}$~GeV and $\tanb = 2$.
The curves correspond to the following transitions:
 $\circ$ into $\nu_{\tau} Z$, $\triangle$ into $\nu_{\tau} J$
 and $\bullet$ into $\tau W$.
The  shaded area will be covered by LEP2.
\end{small}
\end{minipage}

\noindent
\begin{minipage}[t]{156mm}  
{\setlength{\unitlength}{1mm}
\begin{picture}(156,77)
\put(3,2){\mbox{\psfig{figure=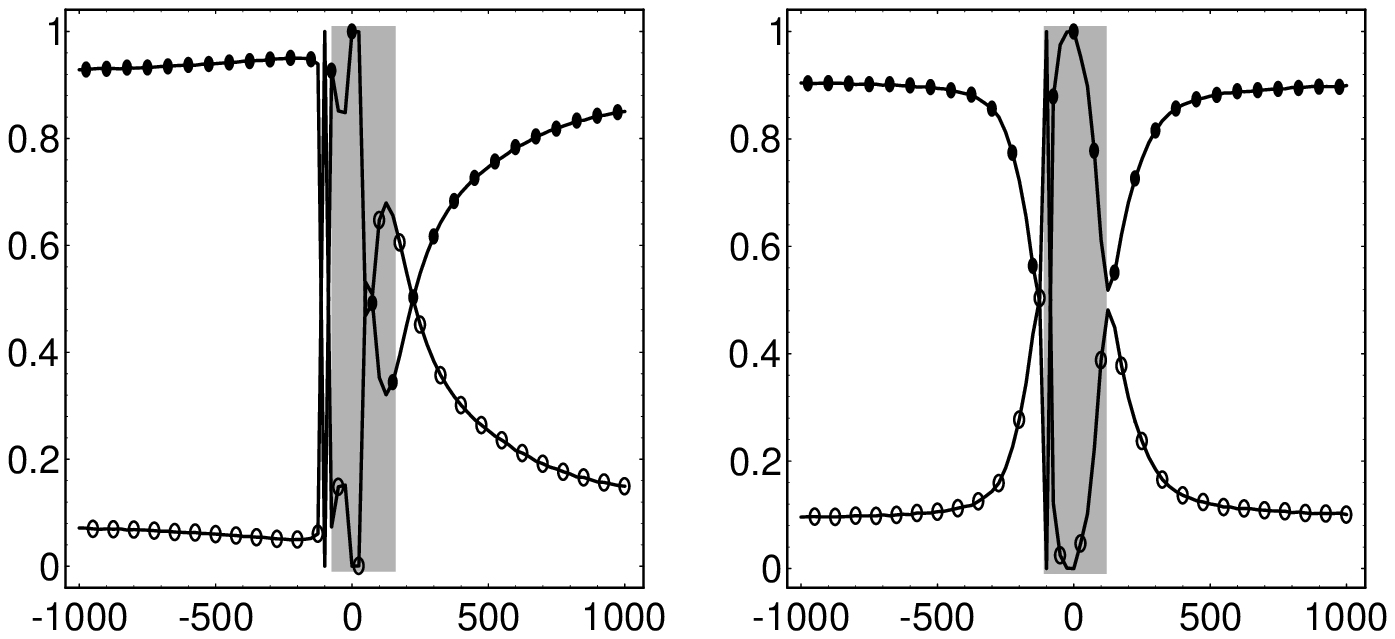,height=6.9cm}}}
\put(3,72){\makebox(0,0)[bl]{{\bf a)}}} 
\put(13.5,69){\makebox(0,0)[bl]{{\small $BR(\chiz{1})$}}}
\put(74, 0.3){\makebox(0,0)[br]{{\small $\mu$~[GeV]}}}
\put(75,72){\makebox(0,0)[bl]{{\bf b)}}} 
\put(85.5,69){\makebox(0,0)[bl]{{\small $BR(\chiz{1})$}}}
\put(146.5, 0.3){\makebox(0,0)[br]{{\small $\mu$~[GeV]}}}
\end{picture}}
\refstepcounter{figure}
\label{brchi01e}
\begin{small}
{\bf \fig{brchi01e}}:
Branching ratios of the lightest neutralino in the $\epsilon$-model. We
have taken $M_2 =170$~GeV, $\epsilon = 1 $~GeV,
a) $\tanb = 2$ and b) $\tanb = 30$.
The curves correspond to the following transitions:
 $\circ$ into $\nu_{\tau} Z$ and $\bullet$ into $\tau W$. 
The  shaded area will be covered by LEP2.
\end{small}
\end{minipage}

\noindent
\begin{minipage}[t]{75mm}  
{\setlength{\unitlength}{1mm}
\begin{picture}(75,77)
\put(3,4){\mbox{\psfig{figure=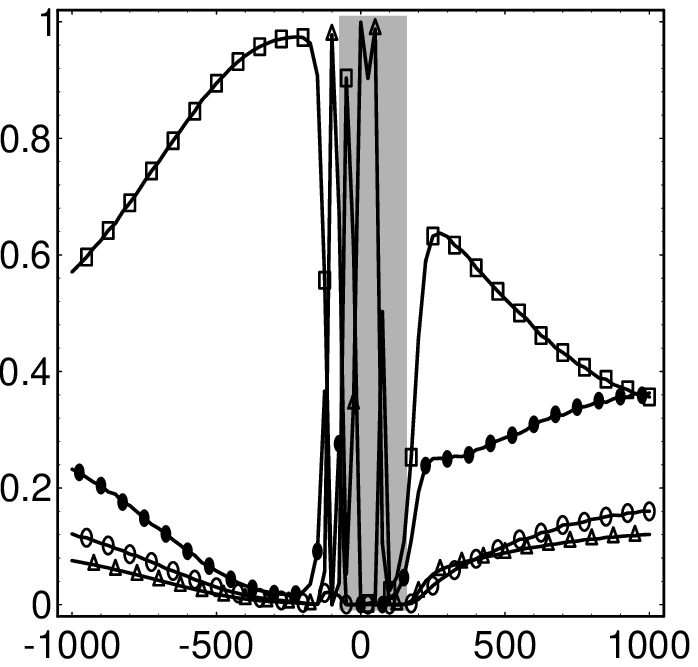,height=6.6cm}}}
\put(8,70){\makebox(0,0)[bl]{{\small $BR(\chiz{2})$}}}
\put(69.5, 0.3){\makebox(0,0)[br]{{\small $\mu$~[GeV]}}}
\end{picture}}
\refstepcounter{figure}
\label{brchi02m}
\begin{small}
{\bf \fig{brchi02m}}:
Branching ratios of the second lightest neutralino in the majoron-model.
We have taken $M_2 =170$~GeV, $h_{\nu33} = 0.01$, $v_R = 100$~GeV,
$v_L = 10^{-5}$~GeV and $\tanb = 2$.
The curves correspond to the following transitions:
 \rechtl into $\chiz{1} Z^{(*)}$,
 $\circ$ into $\nu_{\tau} Z$, $\triangle$ into $\nu_{\tau} J$
 and $\bullet$ into $\tau W$. 
The  shaded area will be covered by LEP2.
\end{small}
\end{minipage}
\hspace{6mm}
\noindent
\begin{minipage}[t]{75mm}  
{\setlength{\unitlength}{1mm}
\begin{picture}(75,77)
\put(3,4){\mbox{\psfig{figure=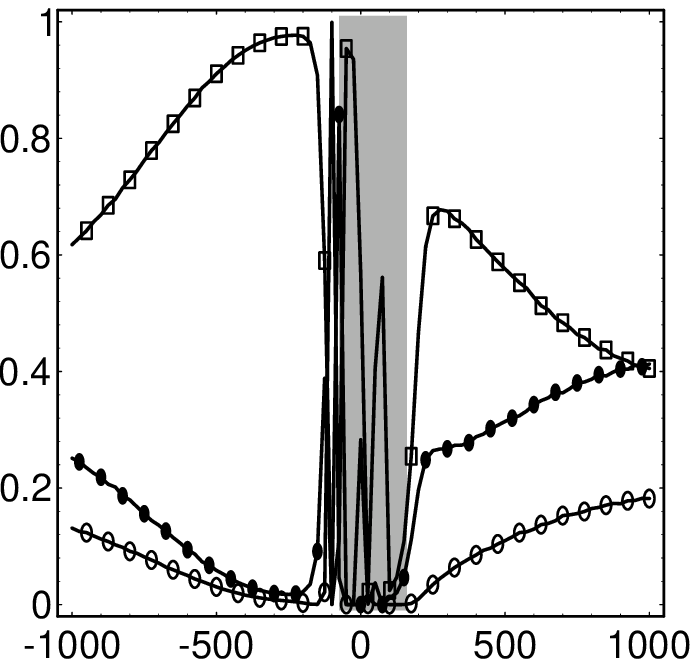,height=6.6cm}}}
\put(8,70){\makebox(0,0)[bl]{{\small $BR(\chiz{2})$}}}
\put(69.5, 0.3){\makebox(0,0)[br]{{\small $\mu$~[GeV]}}}
\end{picture}}
\refstepcounter{figure}
\label{brchi02e}
\begin{small}
{\bf \fig{brchi02e}}:
Branching ratios of the second lightest neutralino in the $\epsilon$-model.
We have taken $M_2 =170$~GeV, $\epsilon = 1 $~GeV and $\tanb = 2$.
The curves correspond to the following transitions:
 \rechtl into $\chiz{1} Z^{(*)}$, $\circ$ into $\nu_{\tau} Z$,
 and $\bullet$ into $\tau W$. The shaded area will be covered by LEP2.
\end{small}
\end{minipage}

\noindent
\begin{minipage}[t]{156mm}  
{\setlength{\unitlength}{1mm}
\begin{picture}(156,77)
\put(3,2){\mbox{\psfig{figure=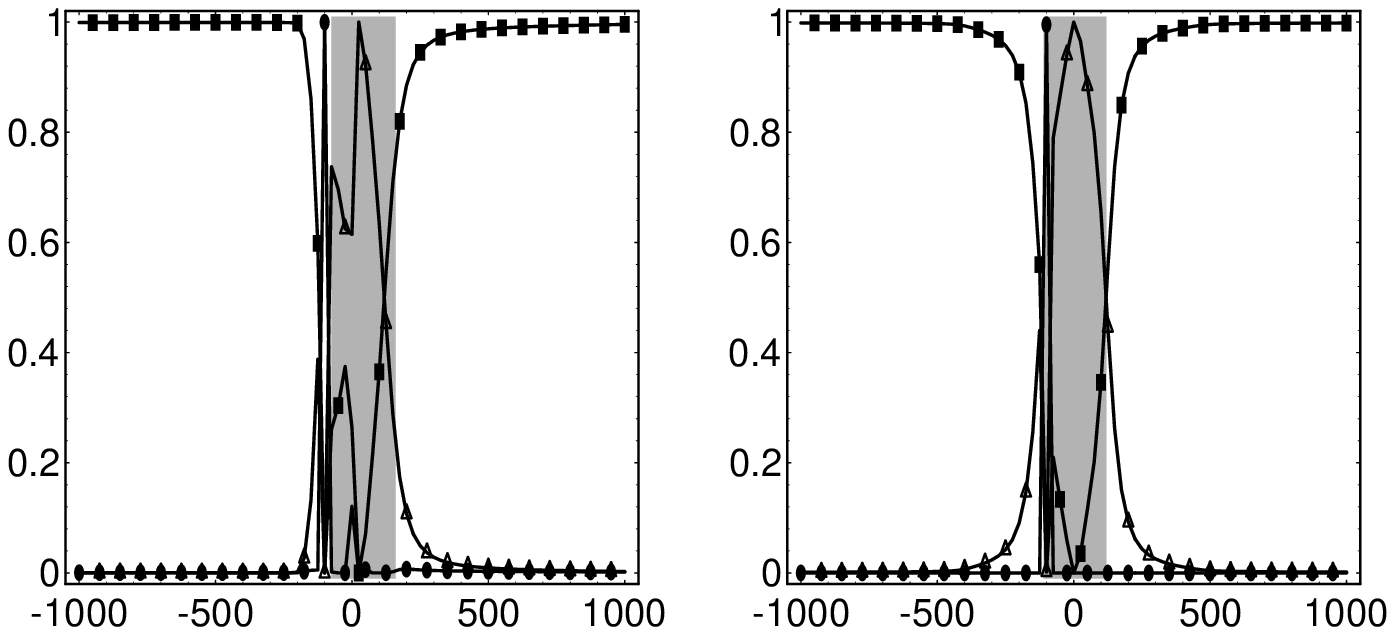,height=6.9cm}}}
\put(3,72){\makebox(0,0)[bl]{{\bf a)}}} 
\put(13.5,69){\makebox(0,0)[bl]{{\small $BR(\chimp{1})$}}}
\put(74, 0.3){\makebox(0,0)[br]{{\small $\mu$~[GeV]}}}
\put(75,72){\makebox(0,0)[bl]{{\bf b)}}} 
\put(85.5,69){\makebox(0,0)[bl]{{\small $BR(\chimp{1})$}}}
\put(146.5, 0.3){\makebox(0,0)[br]{{\small $\mu$~[GeV]}}}
\end{picture}}
\refstepcounter{figure}
\label{brchipm}
\begin{small}
{\bf \fig{brchipm}}:
Branching ratios of the  lightest chargino in the majoron-model. We
have taken $M_2 =170$~GeV, $h_{\nu33} = 0.01$, $v_R = 100$~GeV,
$v_L = 10^{-5}$~GeV,  a) $\tanb = 2$ and b) $\tanb = 30$.
The curves correspond to the following transitions:
 \recht into $\chiz{1} W^{(*)}$, $\triangle$ into $\tau J$
 and $\bullet$ into $\nu_{\tau} W$. 
The  shaded area will be covered by LEP2.
\end{small}
\end{minipage}

\noindent
\begin{minipage}[t]{156mm}  
{\setlength{\unitlength}{1mm}
\begin{picture}(160,170)
\put(3,-22){\mbox{\psfig{figure=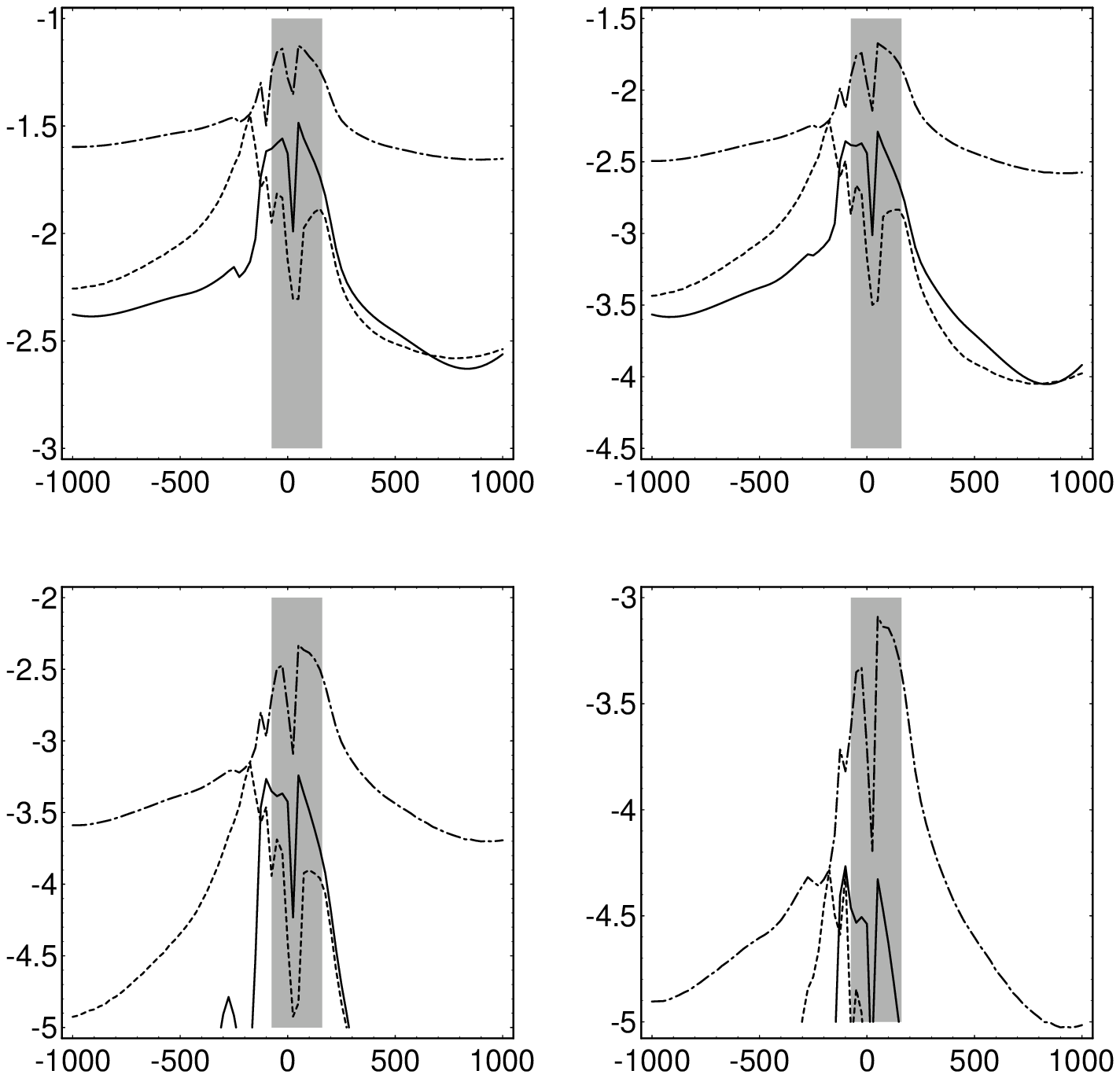,height=19cm}}}
\put(5.5,145){\makebox(0,0)[bl]{{\bf a)}}} 
\put(12.5,143.5){\makebox(0,0)[bl]{
    {\small $\log ( BR(2 \, \glu \to 3 \, l + X))$}}}
\put(75, 74.5){\makebox(0,0)[br]{{\small $\mu$~[GeV]}}}
\put(5.5,70){\makebox(0,0)[bl]{{\bf c)}}} 
\put(12.5,68.5){\makebox(0,0)[bl]{
    {\small $\log ( BR(2 \, \glu \to 5 \, l + X))$}}}
\put(75,-0.3){\makebox(0,0)[br]{{\small $\mu$~[GeV]}}}
\put(80,145){\makebox(0,0)[bl]{{\bf b)}}} 
\put(87,143.5){\makebox(0,0)[bl]{
    {\small $\log ( BR(2 \, \glu \to 4 \, l + X))$}}}
\put(150.0,74.5){\makebox(0,0)[br]{{\small $\mu$~[GeV]}}}
\put(80,70){\makebox(0,0)[bl]{{\bf d)}}} 
\put(87,68.5){\makebox(0,0)[bl]{
    {\small $\log ( BR(2 \, \glu \to 6 \, l + X))$}}}
\put(150.0,-0.3){\makebox(0,0)[br]{{\small $\mu$~[GeV]}}}
\end{picture}}
\refstepcounter{figure}
\label{multilep02}
\begin{small}
{\bf \fig{multilep02}}:
Multi-lepton signals (summed over electrons and muons)
for $\tanb =2$, with other parameters chosen as described before.
We show a) the 3-lepton, b) the 4-lepton, 
c) the 5-lepton and d) the 6-lepton signal for the MSSM (full line), 
the majoron-model (dashed line) and the $\epsilon$-model (dashed-dotted 
line). The  shaded area will be covered by LEP2.
\end{small}
\end{minipage}

\noindent
\begin{minipage}[t]{156mm}  
{\setlength{\unitlength}{1mm}
\begin{picture}(160,170)
\put(3,-22){\mbox{\psfig{figure=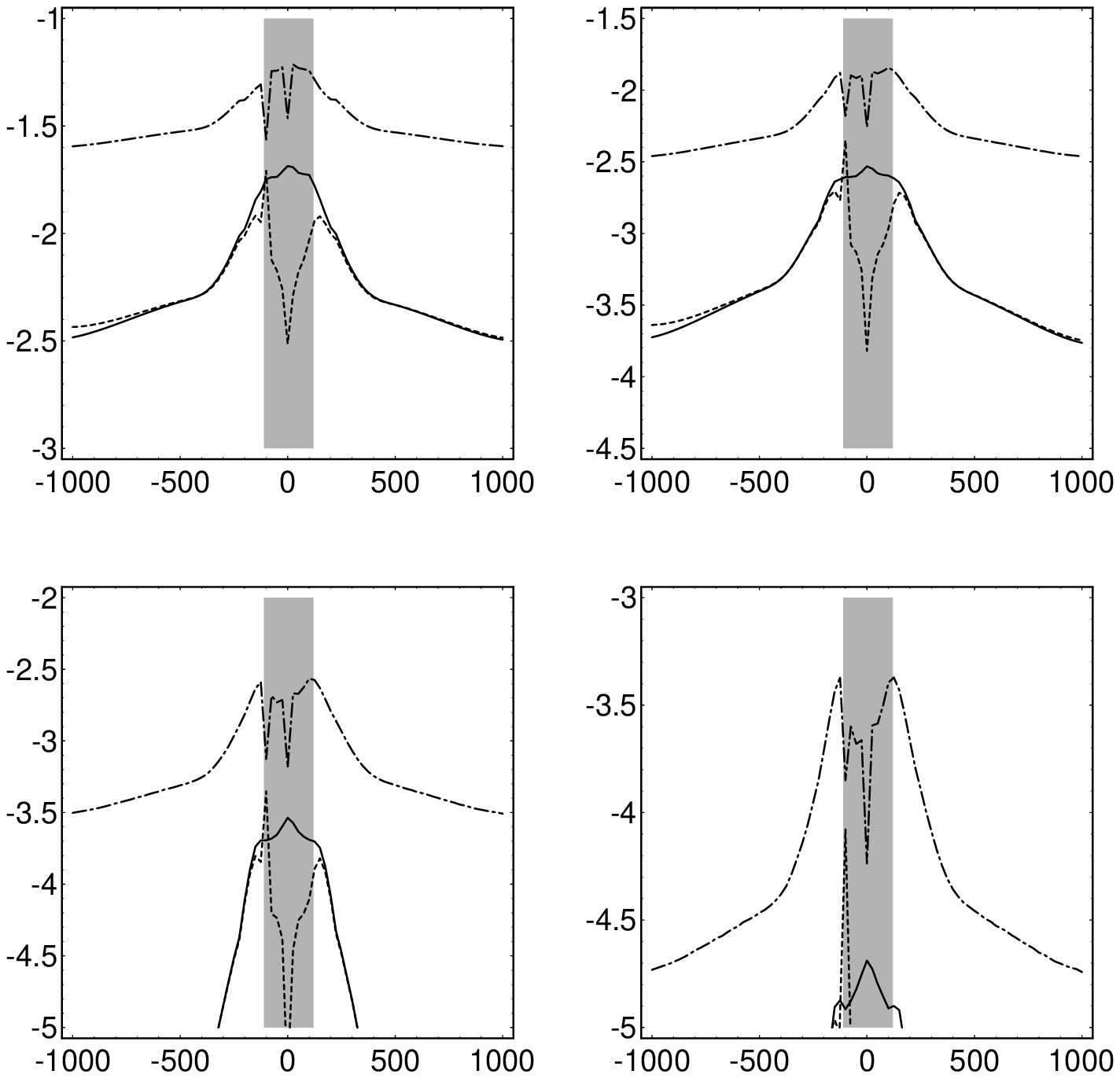,height=19cm}}}
\put(5.5,145){\makebox(0,0)[bl]{{\bf a)}}} 
\put(12.5,143.5){\makebox(0,0)[bl]{
    {\small $\log ( BR(2 \, \glu \to 3 \, l + X))$}}}
\put(75, 74.5){\makebox(0,0)[br]{{\small $\mu$~[GeV]}}}
\put(5.5,70){\makebox(0,0)[bl]{{\bf c)}}} 
\put(12.5,68.5){\makebox(0,0)[bl]{
    {\small $\log ( BR(2 \, \glu \to 5 \, l + X))$}}}
\put(75,-0.3){\makebox(0,0)[br]{{\small $\mu$~[GeV]}}}
\put(80,145){\makebox(0,0)[bl]{{\bf b)}}} 
\put(87,143.5){\makebox(0,0)[bl]{
    {\small $\log ( BR(2 \, \glu \to 4 \, l + X))$}}}
\put(150.0,74.5){\makebox(0,0)[br]{{\small $\mu$~[GeV]}}}
\put(80,70){\makebox(0,0)[bl]{{\bf d)}}} 
\put(87,68.5){\makebox(0,0)[bl]{
    {\small $\log ( BR(2 \, \glu \to 6 \, l + X))$}}}
\put(150.0,-0.3){\makebox(0,0)[br]{{\small $\mu$~[GeV]}}}
\end{picture}}
\refstepcounter{figure}
\label{multilep30}
{\bf \fig{multilep30}}:
\begin{small}
Multi-lepton signals (summed over electrons and muons)
for $\tanb =30$  with other parameters chosen as described before. 
We show a) the 3-lepton, b) the 4-lepton, c) the 5-lepton and d) the 6-lepton
signal for the MSSM (full line), the majoron-model (dashed line)
and the $\epsilon$-model (dashed dotted line).
The  shaded area will be covered by LEP2.
\end{small}
\end{minipage}

\noindent
\begin{minipage}[t]{156mm}  
{\setlength{\unitlength}{1mm}
\begin{picture}(156,77)
\put(3,2){\mbox{\psfig{figure=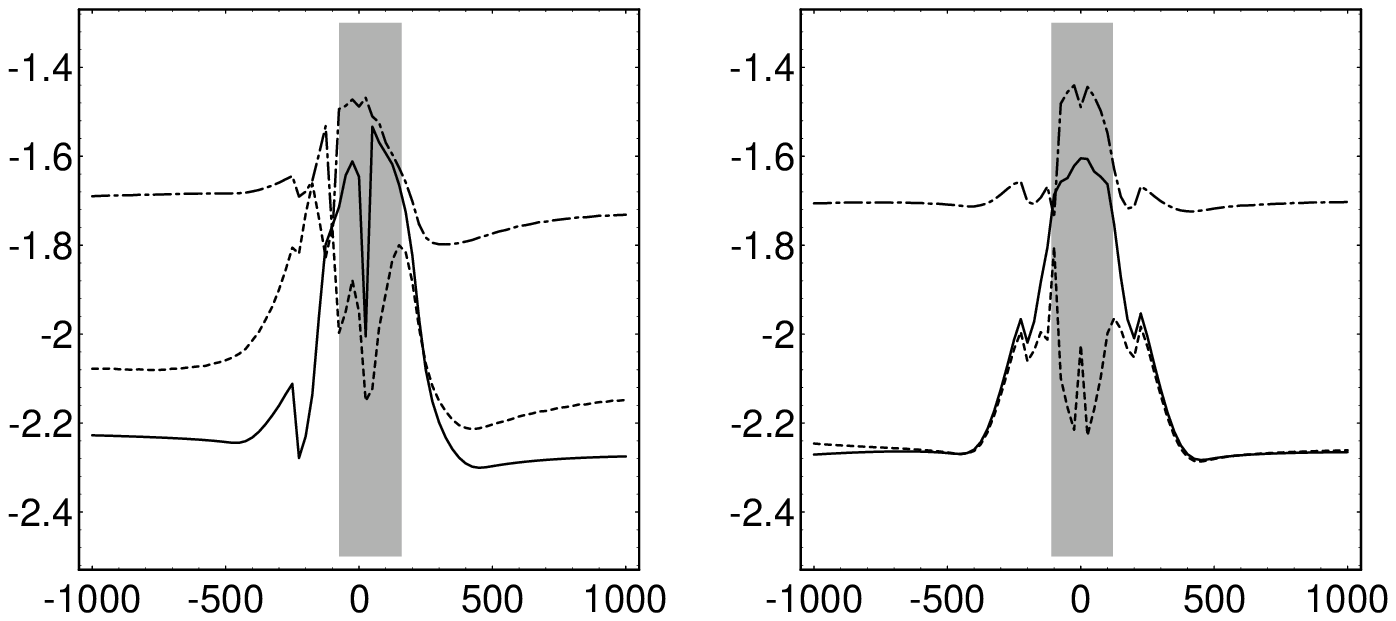,height=6.9cm}}}
\put(3,72){\makebox(0,0)[bl]{{\bf a)}}} 
\put(13.5,69){\makebox(0,0)[bl]{
    {\small $\log ( BR(2 \, \glu \to l^+ l^+, \, l^- l^- ))$}}}
\put(74, 0.3){\makebox(0,0)[br]{{\small $\mu$~[GeV]}}}
\put(75,72){\makebox(0,0)[bl]{{\bf b)}}} 
\put(85.5,69){\makebox(0,0)[bl]{
    {\small $\log ( BR(2 \, \glu \to l^+ l^+, \, l^- l^- ))$}}}
\put(146.5, 0.3){\makebox(0,0)[br]{{\small $\mu$~[GeV]}}}
\end{picture}}
\refstepcounter{figure}
\label{ssdlep}
\begin{small}
{\bf \fig{ssdlep}}:
The same-sign dilepton signal (summed over electrons and muons)
for a) $\tanb =2$ and b) $\tanb = 30$  with other parameters chosen 
as described before. We show situation for the MSSM (full line), the 
majoron-model (dashed line) and the $\epsilon$-model (dashed dotted line).
The  shaded area will be covered by LEP2.
\end{small}
\end{minipage}
\newpage

\begin{appendix}

\section{Squark Couplings}

To get the couplings of the squarks to neutralinos (charginos) and quarks
of \cite{Bartl91} in the notation used in this paper one has to do the
following replacements:
$U_{j1} \to U_{j+3,5}$, $U_{j2} \to U_{j+3,4}$,
$V_{j1} \to V_{j+3,5}$, $V_{j2} \to V_{j+3,4}$,
$N_{j1} \to N_{j+3,7} \cos \theta_W + N_{j+3,6} \sin \theta_W$,
$N_{j2} \to - N_{j+3,7} \sin \theta_W + N_{j+3,6} \cos \theta_W$,
$N_{j3} \to N_{j+3,5} \cos \beta - N_{j+3,4} \sin \beta$ and
$N_{j4} \to N_{j+3,5} \sin \beta + N_{j+3,4} \cos \beta$.

\newpage

\section{Neutralino and Chargino Widths}

Here we collect all the expressions for neutralino and chargino widths.
The decay widths of the two body decays into gauge bosons have the 
generic form
\beq
\Gamma( \tilde{\chi}_i \to \tilde{\chi}_j + V) &=&
 \frac{g^2 \sqrt{\lambda(m^2_i,m^2_j,m^2_V)}}
      {16 \pi \cos^2 \theta_W \, m^3_i}
   \Big[(d^2_L + d^2_R) f_V(m^2_i,m^2_j,m^2_V) \\
      & &  - 6 d_L d_R \epsilon_i \epsilon_j m_i m_j \Big] 
\eeq
with
\beq
\lambda(x,y,z) = (x-y-z)^2 - 4 y z \hspace{2mm}
\eeq
and
\beq
f_V(x,y,z) = \frac{(x-y)^2-2 z^2 + x z + y z}{2z}.
\eeq
The corresponding couplings $d_L$, $d_R$ are given in the table.

For neutralinos, the first three R-parity conserving widths that appear
in (\ref{neutdec}) are given by \cite{bounds}
\beq
\Gamma_{\tilde{\chi}^0_i \to \tilde{\chi}^0_j f\bar{f}} = 
8(v_f^2 + a_f^2)\Gamma^{3b}(M_{\tilde{\chi}^0_i},\tilde{\chi}^0_j,M_Z,
O^{\prime \prime}_{L4j},O^{\prime \prime }_{R4j})
\eeq
and the last three R-parity breaking neutralino widths, which are given  
in \cite{magro}, have the generic form
\beq
\Gamma_{\tilde{\chi}^0_i \to \nu_j f \bar{f}} = 
\Gamma^{3b\prime}(M_i,O^{\prime \prime}_L,O^{\prime \prime}_R,
O^{\prime}_R,O^{\prime}_L,K_L,K_R)
\eeq
For the  Majoron model the width of \eq{majdecay} is given by
\beq
\Gamma_{\tilde{\chi}^0_i \to \tilde{\chi}^0_j J} = 
\frac{1}{16\pi}m_i\left(1-\frac{m_j^2}{m_i^2}\right)O_L^2
\left(1+\frac{m_j^2}{m_i^2}-2\epsilon_i\epsilon_j\frac{m_j}{m_i}\right)
\eeq

For the lightest chargino, we have the following expressions for the
three body decays given in \eq{chardecay2}:
\beq
\Gamma_{\tilde{\chi}^+_1 \to \tilde{\chi}^0_j u\bar{d}} &=&
\Gamma^{3b}(M_{\tilde{\chi}^+},M_{\tilde{\chi}^0_j},M_W,K_{L44},
K_{R44}) \\
\Gamma_{\tilde{\chi}^+_1 \to \tilde{\chi}^0_j \nu_k l^+_k} &=&
\Gamma^{3b\prime\prime}(M_{\tilde{\chi}^+_1},M_{\tilde{\chi}^0_j},M_W,
K_{L4j},K_{R4j})
\eeq
conserving R-parity and
\beq
\Gamma_{\tilde{\chi}^+_1 \to l^+_j f\bar{f}} &=& 
8(v_f^2 + a_f^2)\Gamma^{3b}(M_{\tilde{\chi}^+},0,M_Z,O^{\prime}_{L4j},
O^{\prime}_{R4j})\\
\Gamma_{\tilde{\chi}^+_1 \to \bar{\nu}_j u\bar{d}} &=& 
\Gamma^{3b}(M_{\tilde{\chi}^+},0,M_W,K_{L4j},K_{R4j})\\
\Gamma_{\tilde{\chi}^+_1 \to \bar{\nu}_j \nu_k l^+_k} &=& 
\Gamma^{3b\prime}(M_{\tilde{\chi}^+},K_{L4j},K_{R4j},O^{\prime}_{L4j},
O^{\prime}_{R4j})
\eeq
for the  R-parity-breaking decays. The expressions of $\Gamma^{3b}$ and
$\Gamma^{3b\prime}$ are presented in \cite{bounds}. 
$\Gamma^{3b\prime\prime}$ is given by
\beq
\label{3bnew}
\Gamma^{3b\prime\prime}(m_i,m_j,m_b,d_L,d_R) &=& \frac{e^4 \;m_i}
{256\;\pi^3\;\sin^4{\theta_W}\;\beta^4}\;(d_L^2 + d_R^2)
\;f_1(\beta^2-\delta^2) \nonumber\\
&& + \;2\;d_L \;d_R\;\beta\;\delta \;f_2(\beta^2-\delta^2)
\eeq
with
\beq
f_1(x) &=& -\frac{x^3}{6}-\frac{x^2}{2}+x+(1-x)\ln(1-x) \\
f_2(x) &=& 2x+(2-x)\ln(1-x)
\eeq
and $\beta = \frac{m_i}{m_b}$; $\delta = \frac{m_j}{m_b}$. The
couplings $O^{\prime\prime}$, $O^{\prime}$ and $K$ are the same as in
the table.

In the majoron model, the width for \eq{charmaj} is given by 
\beq
\Gamma_{\tilde{\chi}^+_1 \to l^+_j J} =
\frac{1}{32}m_{\tilde{\chi}^+_1} (C^2_{Lj4} + C^2_{Rj4})
\eeq
with
\beq
\label{cchar}
C_{Lj4}=\frac{v_R}{\sqrt{2}V}\sum^3_{k=1}h_{\nu k3}\eta_4
U_{4k}V_{j4}, \hskip 0.5cm C_{Rj4}=\frac{v_R}{\sqrt{2}V}\sum^3_{k=1}
h_{\nu k3}\eta_j U_{jk}V_{44}\;\;.
\eeq

\begin{table}[tbh]
\label{tabcoup}
\begin{center}
\begin{tabular}{||l|c|l||}
\hline\hline
 & &  \\
$\tilde{\chi}^0_i \to \tilde{\chi}^0_j + Z^0$ & $d_L$ & 
$\frac{1}{\cos \theta_W} O_{Lij}^{\prime\prime}
=  \frac{1}{2 \, \cos \theta_W}\left[ N_{i4} N_{j4} - N_{i5} N_{j5}
- \sum^3_{m=1} N_{im} N_{jm} \right]$ \\ 
& &  \\
\cline{2-3}
& &  \\
 & $d_R$ & $\frac{1}{\cos \theta_W} O_{Rij}^{\prime\prime} = 
- \frac{1}{\cos \theta_W} O_{Lij}^{\prime\prime}$\\ 
& &  \\
\hline
 & &  \\
$\tilde{\chi}^0_i \to \tilde{\chi}^+_j + W^-$ & $d_L$ & $K_{Lji} =
\eta_{j}\left[-\sqrt{2}U_{j5}N_{i6}-U_{j4}N_{i5}-\sum_{m=1}^{3} U_{jm}
N_{im}\right]$ \\
 & &  \\
\cline{2-3}
& &  \\
 & $d_R$ & $K_{Rji} = \epsilon_{i}\left[-\sqrt{2}V_{j5}N_{i6} +
V_{j4}N_{i4}\right]$\\
 & &  \\
\hline
 & &  \\
$\tilde{\chi}^+_i \to \tilde{\chi}^0_j + W^+$ & $d_L$ & $K_{Lij}$ \\
 & &  \\
\cline{2-3}
& &  \\
 & $d_R$ & $K_{Rij}$ \\
 & &  \\
\hline
 & &  \\ 
$\tilde{\chi}^+_i \to \tilde{\chi}^+_j + Z^0$ & $d_L$ &
$O^{\prime}_{Lij} = \frac{1}{2}U_{i4}U_{j4}+U_{i5}U_{j5} + \frac{1}{2}
\sum_{m=1}^{3} U_{im}U_{jm} - \delta_{ij}\sin^2\theta_W$ \\
 & &  \\
\cline{2-3}
& &  \\
 & $d_R$ & $O^{\prime}_{Rij} = \frac{1}{2} V_{i4}V_{j4} + V_{i5}V_{j5}
 - \delta_{ij}\sin^2 \theta_W$\\
 & & \\
\hline\hline
\end{tabular}
\end{center}
\caption{Couplings for neutralino and chargino Charged and Neutral 
Current decays.}
\end{table}

\end{appendix}

\end{document}